\PassOptionsToPackage{draft}{graphicx}
\documentclass[authoryear,preprint,12pt]{elsarticle} 

\setcitestyle{square,numbers,sort&compress,comma}
\usepackage{amsmath}
\usepackage{amssymb}
\usepackage{psfrag}
\usepackage{mhchem}
\usepackage{bm}
\usepackage{flafter}
\usepackage[usestackEOL]{stackengine}

\usepackage{color}
\definecolor{dablue}{rgb}{0,0.4,0.2}
\definecolor{dagr}{rgb}{0,0.5,0.2}
\definecolor{dared}{rgb}{0.4,0,0}
\definecolor{daor}{rgb}{0.7,0.3,0}
\definecolor{gray}{rgb}{0.5,0.5,0.5}
\definecolor{firebrick}{rgb}{0.7,0.13,0.13}
\definecolor{cornflowerblue}{rgb}{0.39,0.58,0.93}
\definecolor{turquoise}{rgb}{0.25,0.88,0.82}
\definecolor{orange}{rgb}{1.0,0.5,0}
\definecolor{purple}{rgb}{0.5,0.0,0.5}

\newcommand{\mi}[1]{{#1}} 
\newcommand{\nn}[1]{{#1}} 
\newcommand{\qm}[1]{{#1}}
\newcommand{\fix}[1]{{#1}}

\newcommand{\opA}{F\mt{_{M}}}
\newcommand{\opB}{F\mt{_{MH}}}
\newcommand{\opC}{F\mt{_{MH}^*}}
\newcommand{\ReacB}{R\mt{_{MH}}}
\newcommand{\ReacC}{R\mt{_{MH}^{*}}}
\newcommand{\ReacD}{R\mt{_{MH}^{0}}}
\newcommand{\RateM}{\qm{$\overline{\dot{\omega}}_{\mathrm{c,CH}_{\mathrm{4}}}$}}
\newcommand{\RateH}{\qm{$\overline{\dot{\omega}}_{\mathrm{c,H}_{\mathrm{2}}}$}}
\newcommand{\PercM}{\qm{$\mathrm{\%\overline{R}_{\methane{}}^{OH}}$}}
\newcommand{\PercH}{\qm{$\mathrm{\%\overline{R}_{\hydrogen{}}^{OH}}$}}
\newcommand{\RM}{\qm{\mt{R_{\methane{}}^{OH}}}}
\newcommand{\RH}{\qm{\mt{R_{\hydrogen{}}^{OH}}}}

\newcommand{\Fref}[1]{Figure~\ref{#1}}
\newcommand{\fref}[1]{Fig.~\ref{#1}}
\newcommand{\methane}{CH\textsubscript{4}}
\newcommand{\hydrogen}{H\textsubscript{2}}
\newcommand{\mt}[1]{$\mathrm{#1}$}
\newcommand{\s}{~}

\newcommand{\sref}[1]{Sec.~\ref{#1}}
\newcommand{\Coh}{\mt{\gamma}\textsubscript{OH}}

\begin{document}

\begin{frontmatter}

\title{\mi{Numerical study of ignition and combustion of \nn{hydrogen-enriched methane} in a sequential combustor}}

\author{Matteo Impagnatiello\corref{cor1}}
\ead{matteoi@ethz.ch}
\author{Quentin Mal\'e\corref{}}
\author{Nicolas Noiray\corref{cor1}}
\ead{noirayn@ethz.ch}
\cortext[cor1]{Corresponding authors}

\address{CAPS Laboratory, Department of Mechanical and Process Engineering, ETH Z\unexpanded{\"u}rich, 8092, Z\unexpanded{\"u}rich, Switzerland}

\begin{abstract}
\mi{Ignition and combustion behavior in the second stage of a sequential combustor are investigated numerically at atmospheric pressure for pure \methane{} fueling and for a \methane{}/\hydrogen{} fuel blend in 24:1 mass ratio using Large Eddy Simulation (LES).
Pure \methane{} fueling results in a turbulent propagating flame anchored by the hot gas recirculation zone developed near the inlet of the sequential combustion chamber. Conversely, \methane{}/\hydrogen{} fueling results in a drastic change of the combustion process, with multiple auto-ignition kernels produced upstream of the main flame \qm{brush}.
Chemical Explosive Mode Analysis indicates that, when \hydrogen{} is added, flame stabilization in the combustion chamber is strongly supported by auto-ignition chemistry. 
The analysis of fuel decomposition pathways highlights that radicals advected from the first stage flame, in particular OH, induce a rapid fuel decomposition and cause the reactivity enhancement that leads to \qm{auto}-ignition upstream of the sequential flame.
This behavior is promoted by the relatively large mass fraction of OH radicals found in the flow reaching the second stage, which is approximately one order of magnitude greater than \qm{it would be at chemical equilibrium}. The importance of the out-of-equilibrium vitiated air on the ignition behavior is proven via an additional LES that features weak auto-ignition kernel formation when equilibrium is artificially imposed.
It is concluded, therefore, that parameters affecting the relaxation towards chemical equilibrium of the vitiated flow can have \qm{an important} influence on the operability of sequential combustors fueled with varying fractions of \hydrogen{} blending.}
\end{abstract}


\end{frontmatter}
\newpageafter{abstract}


\section{Introduction}\label{sec:introduction}
A new type of combustor architecture for heavy duty gas turbines has emerged recently: the Constant Pressure Sequential Combustor (CPSC) \cite{Pennell2017}. This technology is a step forward in the design of operationally- and fuel-flexible \cite{Ciani2019,Guete2008} gas turbines that can cope with both the intrinsically intermittent nature of renewable energy production and with the increasing demand in the power generation sector to burn different fuel blends, and in particular blends incorporating a large amount of sustainably produced hydrogen.
The CPSC concept consists of two technically premixed flames burning in series. No turbine row is positioned between the two stages, differently from previous re-heat designs, making the pressure across the sequential combustor quasi-uniform. In this architecture, \mi{hot flue gases provided by the \nn{lean} first stage flame are diluted with compressor air upstream of the sequential stage.} 
The role of the dilution air is twofold: it increases the O\textsubscript{2} content, allowing for a lean combustion in the sequential stage, and it decreases the gas temperature\mi{, enabling good mixing of the sequential fuel before auto-ignition occurs, thus ensuring very low pollutant emissions.}

The two flames burn at very different thermodynamic conditions, resulting in different combustion modes. The first stage exhibits a purely propagating turbulent flame which is aerodynamically anchored in the combustion chamber as in conventional single stage gas turbine combustors. The reactants are characterized by a relatively low temperature, far below the auto-ignition temperature. In contrast, the temperature of the reactants in the sequential stage is significantly higher, allowing auto-ignition chemistry to play a role in the combustion process \qm{\cite{Schulz2018,Ebi2019,Aditya2019,Gruber:2021,savard}}. 
\fix{In this respect, the competition between diffusion and auto-ignition upstream of the sequential flame can influence the mixture reactivity and flame speed \cite{savard2020}, affecting the combustion process of the second stage.}

Designing combustors operating at auto-ignition conditions requires \fix{therefore} an in-depth understanding of the relevant combustion physics. \fix{Multiple studies have contributed to the problem of auto-ignition in hot turbulent environments in different configurations and with different fuels, e.g. \cite{goussis2015,sidey2014,sidey2015,deng, alex}}. \fix{The combustion behavior of hydrogen at auto-igniting conditions in a sequential combustor configuration has been experimentally investigated in our laboratory in \cite{roberto_asme}. The present study builds on this latter work, but focuses on the effects that methane enrichment with hydrogen has on combustion.}

\fix{The addition of hydrogen to natural gas is promising for sequential combustor architectures, as they were found capable of handling extreme variations of \hydrogen{} content at gas turbine baseload conditions with minor downgrades in \nn{combustor} performances \cite{Bothien2019, Ciani2020,Ciani2021}. In this respect,} there is a need to develop predictive tools for the effect of fuel composition variations, which affect auto-ignition delays, flame dynamics, \nn{and pollutants formation,} \fix{motivating the} research efforts on the topic, e.g., \cite{Fleck:2012,Berger2018}.
In particular, fuel reactivity variations in the sequential stage \fix{of sequential combustors} may increase the NO$_\mathrm{x}$ emissions or damage the burner if the flame anchors at locations that are not designed to sustain high thermal loads \cite{Schmalhofer:2018}.

In this regard, the objective of the present work is to investigate the effects and the challenges of \hydrogen{} blending \nn{on the combustion process} in the second stage of a sequential combustor using numerical tools. Large Eddy Simulation (LES) associated with Analytically Reduced Chemistry (ARC) for the description of the chemical kinetics is used to perform the computations. Chemical Explosive Mode Analysis (CEMA) is used to characterize the combustion regimes. \mi{The influence of radicals advected from the first stage flame on the auto-ignition chemistry is examined via Reaction Path Analysis (RPA), \nn{and} the importance of the relaxation towards chemical equilibrium of the vitiated flow is investigated via an additional, \textit{ad-hoc} LES.}

\section{Computational setup}\label{sec:setup} 

\subsection{Computational domain and boundary conditions}
The configuration investigated in this work is a generic sequential combustor used for research at ETH Z\"urich. Since the physical phenomena of interest occur in the second stage, only this part of the combustor is simulated. 

\begin{figure*}[t!]
\centering
\psfrag{hhhh}[cc][cc][1.0]{\shortstack[c]{\small Dilution Air\\Mixer (DA)}}
\psfrag{sssss}[cc][cc][1.0]{\small Sequential Burner (SB)}
\psfrag{xxxxxxxxxxxxx}[lc][lc][1.0]{\small Combustion Chamber (CC)}

\psfrag{cc}[cc][cc][1.0]{\small 250 mm}
\psfrag{nn}[cc][cc][1.0]{\small 250 mm}

\psfrag{L}[lc][lc][1]{\textcolor{white}{\small L1}}
\psfrag{M}[lc][lc][1]{\textcolor{white}{\small L2}}

\psfrag{aa}[cc][cc][1]{\small Inlet}
\psfrag{bb}[ct][ct][1]{\shortstack[c]{\small Dilution air\\injectors}}

\psfrag{zz}[cc][cc][1]{\small Y\textsubscript{CH\textsubscript{4}} = 0.05}
\psfrag{hh}[cc][cc][1]{\small Heat Release Rate = 10\textsuperscript{8} W/m\textsuperscript{3}}

\psfrag{f}[lc][lc][1]{\small \shortstack[l]{\small Fuel\\injector} }

\psfrag{x}[cc][cc][1]{\small $x$}
\psfrag{y}[cc][cc][1]{\small $y$}
\psfrag{z}[cc][cc][1]{\small $z$}

\psfrag{u}[cc][cc][1]{\footnotesize $x$}
\psfrag{r}[cc][cc][1]{\footnotesize $y$}
\psfrag{v}[cc][cc][1]{\footnotesize $z$}

\psfrag{oo}[cc][cc][1]{\small Outlet}

\includegraphics[width=\textwidth, draft=False]{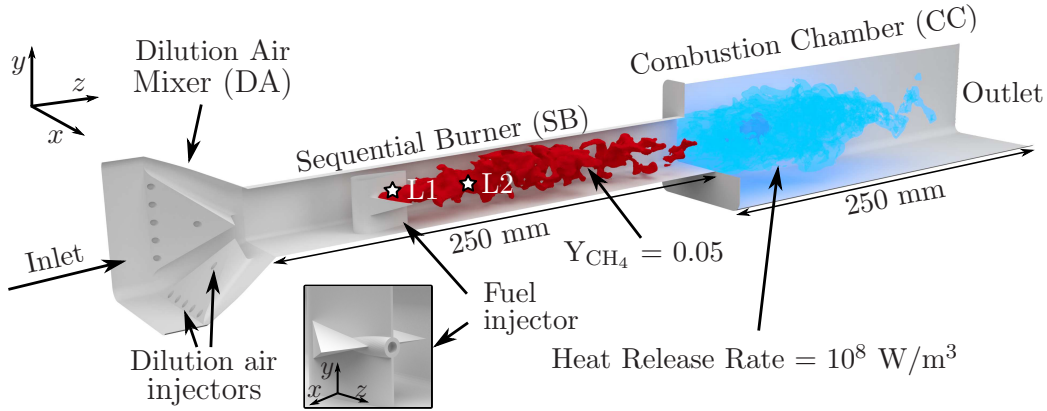} 
\caption{Geometry of the second stage of the sequential combustor used for the 3D LES computations. L1 and L2 identify the locations that will be considered in the reaction path analysis.}
\label{fig:rendering}
\end{figure*}

The simulated sequential stage is shown in Fig.~\ref{fig:rendering} and consists of three parts: the Dilution Air mixer (DA), the Sequential Burner (SB) and the Combustion Chamber (CC). The DA module connects the first stage combustion chamber with the SB. It is equipped with large lateral vortex generators and multiple air injection holes for rapid mixing. The axial fuel injector is located inside the SB. Its shape is an extruded, symmetric airfoil with two vortex generators attached on each side to favor mixing between the fuel and the co-flow of hot vitiated \qm{air} \nn{flow}.

\setstackgap{S}{3pt}

\mi{In the present configuration, the inlet mixture consists of 15.7 g/s of combustion products from a \nn{lean} premixed methane/air flame burning at an equivalence ratio of 0.8 in the first stage. The inlet temperature is set to 1860~K, \nn{which corresponds to} 93\% of the adiabatic flame temperature, in order to take into account the heat losses at the walls of the first stage combustion chamber. Chemical equilibrium at 1860~K is assumed.}

The \mi{inlet mixture} is diluted in the DA mixer with 8.5 g/s of air at ambient temperature, leading to a lean combustion in the sequential stage.
The fuel mass flow in the sequential stage is 0.6 g/s. Throughout this study, all the parameters discussed above are fixed. Two cases are simulated, which differ only in terms of sequential fuel composition:
\begin{itemize}
    \item[-] case \opA{} where the fuel is pure \methane;
    \item[-] case \opB{} where the fuel is a mixture of CH\textsubscript{4}:H\textsubscript{2} in 24:1 mass ratio, which corresponds to a volume fraction of H$_2$ of 25\%.
\end{itemize}
The equivalence ratios are 0.9 and 0.94 for \opA{} and \opB{}, respectively. The lateral boundary condition of the simulated domain is defined as an isothermal condition. The temperature of the DA mixer boundary is set to 720 K, the one of the fuel injector to 420 K and the one of the SB to 1020 K. These temperatures are based on simplified conjugate heat transfer simulations at nominal operating conditions, which were performed during the conception of the experimental sequential combustor and included the specific water and air cooling systems of these components.

\noindent The simulations are performed at atmospheric pressure.

\subsection{Large eddy simulation setup}
LES are performed using the explicit massively parallel compressible solver AVBP \cite{Schonfeld1999,Gicquel2011}. \mi{The numerical framework used in this work has been experimentally validated in \qm{\cite{Schulz2019, male2022}} for a similar sequential combustor \nn{geometry}. The \nn{present} LES results are therefore considered a trustworthy database for the type of analysis and the purpose of this work.}

A fully explicit two-step Taylor–Galerkin finite element numerical scheme \cite{Colin2000a} is used, which offers third order accuracy in space and time. The sub-grid scale turbulence effect is modeled using the SIGMA model \cite{Nicoud2011}. Classical law-of-the-wall modeling is used to account for the unresolved boundary layer \cite{Vandriest}. The Navier–Stokes Characteristic Boundary Conditions (NSCBC) formulation \cite{Poinsot1992} is applied to the inlets and the outlet of the domain. The Dynamic Thickened Flame (DTF) model is used to model turbulent combustion \cite{Legier2000}. Unresolved flame wrinkling effects are modeled using the efficiency function defined in Ref.~\cite{Charlette2002}.
To ensure that the auto-ignition is not biased by the DTF model in the sequential burner \cite{Schulz2017,Schulz2019d}, the activation of the model is limited to the combustion chamber only.

\begin{figure}[t!]
    \centering
    \psfrag{a}[lb][lb][1.0]{a)}
    \psfrag{aa}[lb][lb][1.0]{b)}
    \psfrag{aaa}[lb][lb][1.0]{c)}
    \psfrag{GRI}[tr][tr][1.0]{GRI}
    \psfrag{IGR}[tr][tr][1.0]{ARC}
    \psfrag{Linse}[tl][tl][1.0]{Symbols:}
    \psfrag{Lines}[tl][tl][1.0]{Lines:}
    \psfrag{0D}[tc][tc][1.0]{\qm{0D Reactors}}
    \psfrag{D0}[tc][tc][1.0]{\qm{1D Flame}}

    \psfrag{z}[tc][tc][1.0]{$Z$ [-]}
    \psfrag{tau}[][][1.0]{$\tau_{AI}$ [ms]}
    \psfrag{20}[rc][rc][1.0]{100}
    \psfrag{10}[rc][rc][1.0]{10}
    \psfrag{00}[rc][rc][1.0]{1}
    \psfrag{0.001}[cc][cc][1.0]{0.001}
    \psfrag{0.01}[cc][cc][1.0]{0.01}
    \psfrag{Fm}[][][1.0]{\opA}
    \psfrag{Fmh}[][][1.0]{\textcolor{gray}{\opB}}
    \psfrag{pw}[cb][cc][1.0]{$\phi_{\omega}$}
    \psfrag{ps}[][][1.0]{\textcolor{cornflowerblue}{$\phi_{s}$}}
    \psfrag{la}[][][1.0]{\textcolor{firebrick}{$\lambda_e$}}
    \psfrag{CH2O}[][][1.0]{\textcolor{firebrick}{CH\textsubscript{2}O}}
    \psfrag{T}[][][1.0]{T}
    \psfrag{CH4}[][][1.0]{\textcolor{cornflowerblue}{CH\textsubscript{4}}}
    \psfrag{x-xx}[tc][tc][1.0]{$x-x_f$ [mm]}
    \psfrag{EI}[][][1.0]{EI}
    \psfrag{phi}[][][1.0]{$\phi/\phi_{\omega, max}$}
    \psfrag{lam}[][][1.0]{\textcolor{firebrick}{$\log_{10}(1+\lambda)$}}
    \psfrag{-1}[cc][cc][1.0]{-1}
    \psfrag{0}[cc][cc][1.0]{0}
    \psfrag{1}[lc][lc][1.0]{\textcolor{firebrick}{1}}
    \psfrag{i}[rc][rc][1.0]{0}
    \psfrag{l}[rc][rc][1.0]{1}
    \psfrag{3}[lc][lc][1.0]{\textcolor{firebrick}{3}}
    \psfrag{5}[lc][lc][1.0]{\textcolor{firebrick}{5}}
    \psfrag{0.3}[lc][lc][1.0]{0.3}
    \psfrag{0.6}[lc][lc][1.0]{0.6}
    \psfrag{0.0}[lc][lc][1.0]{0}
    \includegraphics[width=\textwidth, draft=False]{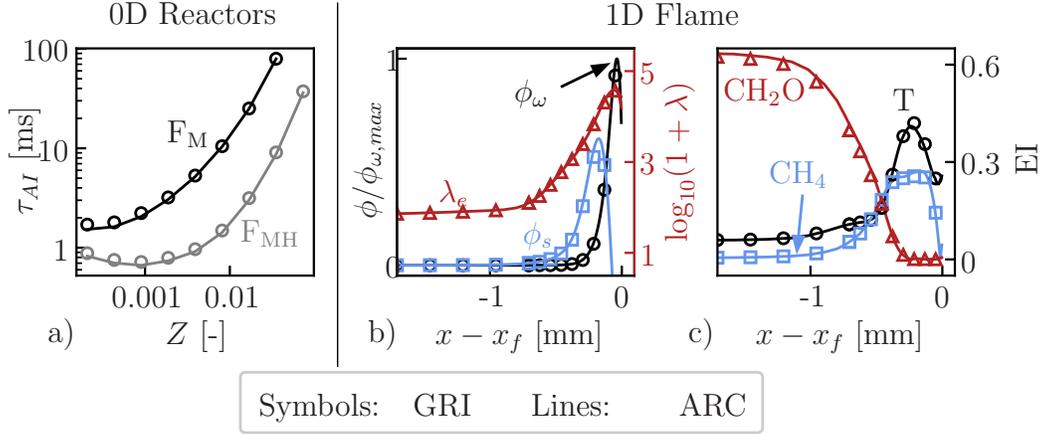}
    \caption{\qm{Comparison of the detailed (GRI) and the ARC mechanisms in terms of auto-ignition time $\tau_{AI}$ (a) and in terms of important CEMA-related quantities introduced in \sref{sec:CEMA} (b and c). $x_f$ represents the flame front location.}}
    \label{fig:mechanism}
\end{figure}

The chemical kinetics relies on ARC. The ARC mechanism consists of 19 transported species, 8 species in quasi-steady state approximation and 142 reactions. This scheme has been reduced from the GRI3.0 mechanism \cite{gri30} using the YARC tool \cite{Pepiot-Desjardins2008,Pepiot:phd2008}\mi{, and has been validated via 0D and 1D Cantera \cite{cantera} simulations.} A very good agreement in terms of both ignition delay and laminar flame speed has been found between the detailed and the reduced scheme. Ignition delays as function of the mixture fraction \qm{$Z$} between the vitiated air and the second stage fuel are shown in \fref{fig:mechanism}a for both \opA{} and \opB{}. \mi{Additionally, the reduced and the detailed schemes showed great agreement in terms of CEMA-related quantities (introduced in \sref{sec:CEMA}). The validation was performed by means of 1D flame simulations at multiple operating conditions representative of the ones found in the 3D LES. The results for some of the most important CEMA-related quantities are illustrated in \fref{fig:mechanism}b \nn{and \ref{fig:mechanism}c}, in one of the tested operating points as function of the streamwise coordinate.}

\begin{figure*}[t!]
\centering
\psfrag{A}[tc][tc][1.0]{0.65}
\psfrag{B}[tc][tc][1.0]{0.80}
\psfrag{C}[tc][tc][1.0]{0.95}
\psfrag{D}[bc][bc][1.0]{0.5}
\psfrag{E}[bc][bc][1.0]{1.2}
\psfrag{S}[tc][tc][1.0]{LESIQ\textsubscript{$\nu$} [-]}
\psfrag{R}[bc][bc][1.0]{Mesh size [mm]}
\psfrag{y}[cc][cc][1.0]{$y$}
\psfrag{z}[cc][cc][1.0]{$z$}
\includegraphics[width=\textwidth, draft=False]{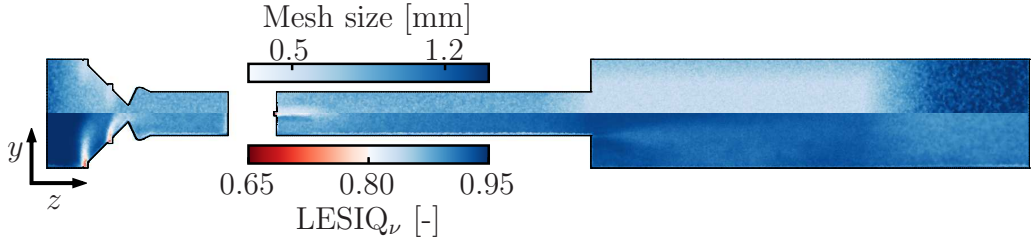}
\caption{Planar cut colored by the mesh size (top half) and the LESIQ\textsubscript{$\nu$} criterion (bottom half).
}
\label{fig:mesh_lesiq}
\end{figure*}

The computational mesh is unstructured and comprises 24 million tetrahedral elements. The characteristic size of each element is \mt{0.5\s mm} in the flame region and close to the injectors, and \mt{0.75\s mm} otherwise, as shown in \fref{fig:mesh_lesiq}. The mesh size slowly increases to \mt{1.2\s mm} at the inlet and at the outlet of the domain. Grid quality is assessed using the LES Index of Quality criterion based on turbulent viscosity (LESIQ\textsubscript{$\nu$}) \cite{Celik2005}. A non-reactive LES returned a value of LESIQ\textsubscript{$\nu$} between 0.8 and 0.95 (\fref{fig:mesh_lesiq}). This highlights a fair resolution of the turbulent spectrum in the LES framework. The ratio between the mesh size and the laminar flame thickness is about 1.4 in the CC. The Thickening Factor (TF) is adjusted so that 4.5 grid points are located inside the flame front, resulting in an applied TF of about~6.

\begin{figure*}[t!]
\centering
\psfrag{Z}[rc][rc][1.0]{\small Z1}
\psfrag{ZZ}[rc][rc][1.0]{\small Z2}

\psfrag{A}[tc][tc][1.0]{-50}
\psfrag{B}[tc][tc][1.0]{0}
\psfrag{G}[tc][tc][1.0]{100}
\psfrag{I}[tc][tc][1.0]{200}
\psfrag{D}[bc][bc][1.0]{600}
\psfrag{E}[bc][bc][1.0]{1300}
\psfrag{F}[bc][bc][1.0]{2000}
\psfrag{S}[tc][tc][1.0]{w [m/s]}
\psfrag{R}[bc][bc][1.0]{Temperature [K]}
\psfrag{y}[cc][cc][1.0]{$y$}
\psfrag{z}[cc][cc][1.0]{$z$}
\includegraphics[width=\textwidth, draft=False]{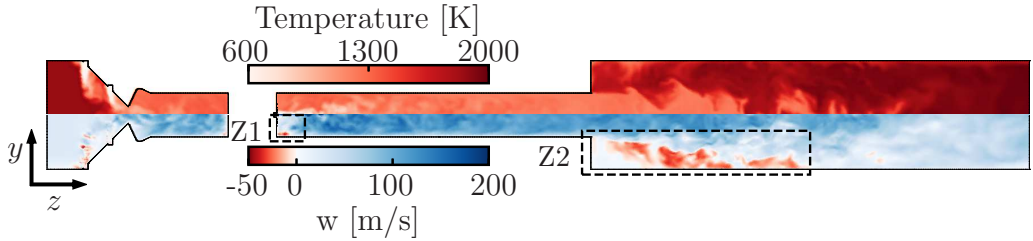}
\caption{Planar cut colored by the temperature (top half) and the streamwise velocity (bottom half). The recirculation zones Z1 and Z2 are highlighted by the dashed rectangles.}
\label{fig:TV}
\end{figure*}

\Fref{fig:TV} provides the instantaneous fields of the temperature and of the axial velocity obtained from LES under \opA{} conditions.
The DA mixer enables rapid decrease of the temperature before the sequential fuel injector. The presence of  hot temperature spots at this location is thereby minimized. Despite the high turbulence intensity produced by the vortex generators, temperature and composition stratification persist in the SB for the present configuration. Fuel mixing with the co-flow occurs through the whole length of the SB. The streamwise velocity field highlights the presence of two \nn{critical} recirculation zones. The first one \mi{(Z1)}, located right after the fuel injector, supports the formation of a chemically-active region that promotes fuel decomposition \mi{and that} will be scrutinized  in \sref{sec:ignitionmechanism}. The second recirculation zone \mi{(Z2)} is located in the CC and strengthens the flame stabilization. Hot combustion products and radicals are trapped in this recirculation zone and support flame anchoring at the burner outlet, thereby inhibiting sudden flame blow-off when the ignition delay becomes significantly longer than the residence time in the SB \nn{\cite{SchulzNoiray2019, Ebi2019}}.

\subsection{Chemical Explosive Mode Analysis}\label{sec:CEMA}
CEMA is a diagnostic tool for reacting flows. It was introduced by Lu et al. \cite{Lu2010} based on concepts from Computational Singular Perturbation (CSP) \cite{Lam}. The principle of CEMA can be formalized from the differential equation that governs the evolution of the thermochemical state of a reacting flow:
\begin{equation}
    \frac{D\mathbf{y}}{Dt}=\mathbf{g}(\mathbf{y}) = \bm{\omega}(\mathbf{y}) + \mathbf{s}(\mathbf{y})  \text{,}
\end{equation}
with $D\mathbf{y}/Dt$ being the material derivative of the thermochemical state vector $\mathbf{y}$, and $\bm{\omega}$ and $\mathbf{s}$ being the chemical and the diffusion source term vectors, respectively.
CEMA focuses only on the chemical kinetics, looking at the linear stability of a particular thermochemical state through the eigenanalysis of the local chemical Jacobian matrix $\bm{J_\omega}$.

Among the different modes of $\bm{J_\omega}$, CEMA focuses on the Chemical Explosive Mode (CEM), namely a mode whose eigenvalue has a  positive real part. The existence of a CEM for a given mixture shows the tendency of that specific mixture to ignite: an infinitesimal perturbation of that mode would exponentially grow in an isolated environment. If multiple modes have positive eigenvalues, CEM traditionally refers to the fastest mode, namely the one with the largest eigenvalue $\lambda_e$.

The Explosive Index $\mathbf{EI}$ was introduced as an algorithmic tool for the identification of variables \nn{having the greatest influence on} the CEM. It is defined as follows:
\begin{equation}
    \mathbf{EI} = \frac{\lvert \mathrm{diag}(\mathbf{a_e}\mathbf{b_e})\rvert}{\mathrm{sum}(\lvert \mathrm{diag}(\mathbf{a_e}\mathbf{b_e})\rvert)}  \, \text{,}
\end{equation}
where $\mathbf{a_e}$ and $\mathbf{b_e}$ are the right and left eigenvectors associated with $\lambda_e$, respectively.

The local combustion mode indicator \mt{\alpha} has been introduced by Xu et al. \cite{Xu2019}. It compares the local diffusion and chemical source terms after projecting them along the CEM. Therefore, \mt{\alpha} is defined only in the regions where a CEM exists and it is computed as follows:
\begin{equation}
    \alpha = \frac{\mathbf{b_e}\cdot\mathbf{s}}{\mathbf{b_e}\cdot\bm{\omega}} = \frac{\phi_s}{\phi_\omega} \, \text{,}
\end{equation}
where \mt{\phi_s} and \mt{\phi_\omega} are the projections of the diffusion and chemical source terms on the left eigenvector $\mathbf{b_e}$, respectively.
Depending on the value of \mt{\alpha}, three flame regimes can be identified: (i) propagation-dominated regime where the diffusion term dominates the chemical term and promotes ignition (\mt{\alpha>1}); (ii) extinction-dominated regime where the diffusion term dominates the chemical term but acts against ignition (\mt{\alpha<-1}); and (iii) auto-ignition dominated regime, where the chemical term dominates the diffusion term (\mt{\lvert \alpha \rvert<1}).

In this study, CEMA is used to distinguish the regions where the mixture is explosive, characterise the local combustion regime and identify the variables related the most to the CEM. \nn{This method has already been used to successfully characterize the combustion regime of the present sequential combustor in \cite{male2022}.}

\begin{figure*}[t!]
\centering
\psfrag{EEE}[bc][bc][1.0]{10\textsuperscript{7}}
\psfrag{FFF}[bc][bc][1.0]{10\textsuperscript{8}}
\psfrag{GGG}[bc][bc][1.0]{10\textsuperscript{9}}
\psfrag{aa}[lc][lc][1.0]{$t$}
\psfrag{ee}[lc][lc][1.0]{$t+\Delta t$}
\psfrag{y}[cc][cc][1.0]{$y$}
\psfrag{z}[cc][cc][1.0]{$z$}
\psfrag{RR}[cc][cc][1.0]{\opA}
\psfrag{PP}[cc][cc][1.0]{\opA}
\psfrag{NN}[cc][cc][1.0]{\opB}
\psfrag{HH}[cc][cc][1.0]{\opB}
\psfrag{TF}[lc][lc][1.0]{TF [-]}
\psfrag{1}[rc][rc][1.0]{1}
\psfrag{10}[rc][rc][1.0]{10}
\psfrag{HRR}[cb][cb][1.0]{$\dot{Q}$ [W/m\textsuperscript{3}]}
\includegraphics[width=\textwidth, draft=False]{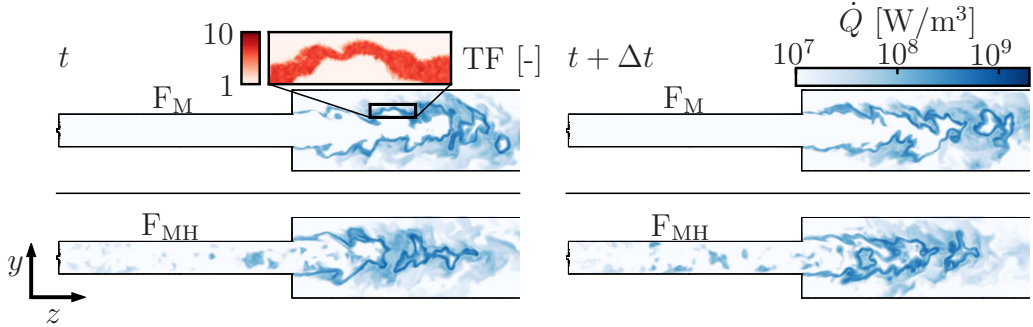}
\caption{Planar cuts colored by $\dot{Q}$ for both \opA{} (top) and \opB{} (bottom). Two different instants separated by $\Delta t=0.6~\mathrm{ms}$ are represented, after a statistically steady state has been reached. The inset shows the Thickening Factor (TF) field.}
\label{fig:hr}
\end{figure*}

\begin{figure*}[t!]
    \centering
    \psfrag{XX}[cc][cc][1.0]{\textbf{Alpha}}
    \psfrag{aa}[rt][rc][1.0]{a)}
    \psfrag{ee}[rt][rc][1.0]{b)}
    \psfrag{RR}[cc][cc][1.0]{\opA}
    \psfrag{PP}[cc][cc][1.0]{\opB}
    \psfrag{Al}[tc][tc][1.0]{\mt{-\infty}}
    \psfrag{BB}[t][t][1.0]{\mt{-1}}
    \psfrag{CC}[t][t][1.0]{\mt{1}}
    \psfrag{Dl}[tc][tc][1.0]{\mt{+\infty}}
    \psfrag{alpha}[cc][cc][1.0]{\mt{\alpha}}
    \psfrag{EI}[cc][cc][1.0]{EI}
    \psfrag{LL}[cc][cc][1.0]{\textbf{Explosive Index}}
    \psfrag{NN}[cc][cc][1.0]{\opA}
    \psfrag{HH}[cc][cc][1.0]{\opB}
    \psfrag{EEE}[t][t][1.0]{\small T}
    \psfrag{FFF}[t][t][1.0]{\small CH\textsubscript{3}}
    \psfrag{GGG}[t][t][1.0]{\small CH\textsubscript{2}O}
    \psfrag{HHH}[t][t][1.0]{\small Other}
    \psfrag{y}[][][1.0]{$y$}
    \psfrag{z}[][][1.0]{$z$}
    \includegraphics[width=\textwidth, draft=False]{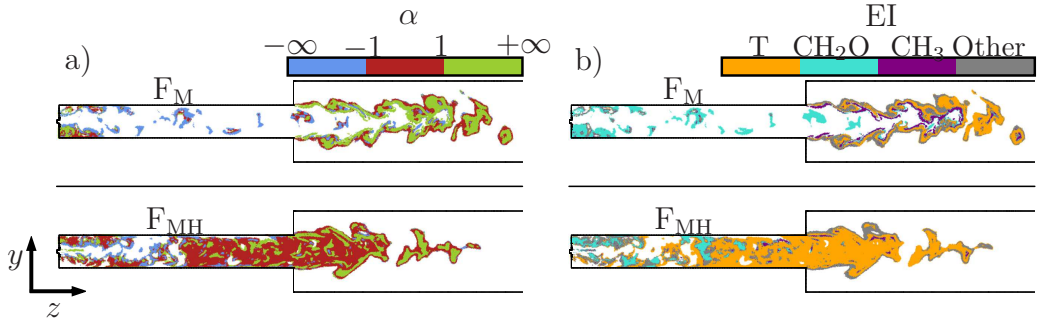}
\caption{Planar cuts colored by \mt{\alpha} (a) and \bm{\mt{EI}} largest entry (b), for both \opA{} (top) and \opB{} (bottom).}
\label{fig:CEMA}
\end{figure*}

\section{Numerical results}

\subsection{Combustion regimes}
The \nn{heat release rate ($\dot{Q}$)} field on a central plane of the domain is shown in \fref{fig:hr} for both operating points. 
\hydrogen{} addition affects the chemical activity in the SB and the shape of the flame in the CC. Multiple auto-ignition kernels are continuously initiated in the SB and advected downstream where they merge with the flame front in the CC.
This is not seen in \opA, where combustion only occurs in the CC. From these observations, it can be presumed that in \opA{} the flame combustion regime is dominated by propagation through diffusion of heat and radicals, while in \opB{} the auto-ignition chemistry plays an important role and has a strong effect on the stabilization of the flame in the CC.

CEMA is used to precisely characterize the combustion mechanisms. The combustion mode indicator \mt{\alpha} and the largest \mt{\mathbf{EI}} entries are shown in \fref{fig:CEMA} on a central plane of the domain for explosive mixtures. A threshold value of $\lambda_e = 100~\mathrm{s^{-1}}$ is used to \mi{take into consideration only the CEMs whose characteristic times are lower or comparable to the sequential burner residence time. Only a single instant is represented in \fref{fig:CEMA} since the variations in the \mt{\alpha} and \mt{\mathbf{EI}} fields are due to the unsteady nature of the turbulent flow, making the observations independent of the specific snapshot considered.}

The contours of $\alpha$ clearly show the change in combustion regime. For \opA, the CC flame features a \mi{clear} premixed propagating flame structure with a pre-heat zone where \mt{\alpha>1} followed by a thin reaction zone where \mt{\lvert \alpha \rvert<1}. For \opB, on the contrary, \mt{\alpha} exhibits large \mt{\lvert \alpha \rvert<1} regions \mi{upstream of the \qm{sequential flame} highlighting that, in this case, the flame in the CC is not purely propagating, but supported by auto-ignition chemistry.} 

Chemical activity near the fuel injector body is \nn{observed from} CEMA for both cases. The largest \mt{\mathbf{EI}} entry is mainly associated with CH\textsubscript{2}O, an intermediate oxidized product of \methane{}. This shows that some \methane{} oxidation already occurs in this region. Further down in the SB, almost no fast enough CEM (i.e., $\lambda_e \geq 100 \, \mathrm{s^{-1}}$) is present for the \opA{} case. On the contrary, for \opB{}, fast CEMs are present due to the auto-igniting tendency of the advected mixture. In this case the largest \mt{\mathbf{EI}} entry is associated with temperature, indicating that the explosive mixture is going through thermal runaway. The chemical activity in the immediate vicinity of the fuel injector body revealed by CEMA is a-priori unexpected. The chemical mechanisms involved are investigated in more detail in the following sections.

\subsection{Fuel oxidation mechanisms}
\label{sec:ignitionmechanism}
Reaction Path Analysis (RPA) is used to investigate the mechanisms leading to the oxidation of the fuel in the SB. The analysis is performed using thermochemical conditions representative of \mi{the locations L1 and L2 indicated in \fref{fig:rendering}. The former is \nn{located} in the recirculation zone near the fuel injector where the fuel mixes with the vitiated air stream; the latter is located downstream of the fuel injector.}

\begin{figure*}[t!]
\centering
\psfrag{l1}[rc][rc][1.0]{\small L1}
\psfrag{l2}[rc][rc][1.0]{\small L2}
\psfrag{fm}[bc][bc][1.0]{\small \opA{}}
\psfrag{fmh}[bc][bc][1.0]{\small \opB{}}

\setstackgap{S}{0pt}
\psfrag{aa}[lt][lt][0.8]{\Shortstack[l]{a)\\\small Scale = 1}}
\psfrag{bb}[lt][lt][0.8]{\Shortstack[l]{b)\\\small Scale = 1}}
\psfrag{cc}[lt][lt][0.8]{\Shortstack[l]{c)\\\small Scale=10\textsuperscript{4}}}
\psfrag{dd}[lt][lt][0.8]{\Shortstack[l]{d)\\\small Scale=10\textsuperscript{2}}}

\psfrag{CHl}[cc][cc][0.8]{\footnotesize CH\textsubscript{4}}
\psfrag{Hl}[cc][cc][0.8]{\footnotesize H\textsubscript{2}}
\psfrag{ClH}[cc][cc][0.8]{\footnotesize CH\textsubscript{3}}
\psfrag{HlO}[cc][cc][0.8]{\footnotesize H\textsubscript{2}O}
\psfrag{H}[cc][cc][0.8]{\footnotesize H}
\psfrag{OH}[cc][cc][0.8]{\footnotesize OH}
\psfrag{ClHl}[cc][cc][0.8]{\footnotesize C\textsubscript{2}H\textsubscript{6}}
\psfrag{CHlO}[cc][cc][0.8]{\footnotesize CH\textsubscript{3}O}
\psfrag{ClHO}[cc][cc][0.8]{\footnotesize CH\textsubscript{2}O}

\setstackgap{S}{0.5pt}
\psfrag{47}[ct][ct][0.8]{\footnotesize 4.7}
\psfrag{13}[ct][ct][0.8]{\footnotesize 1.3}
\psfrag{a}[rc][rc][0.8]{\Shortstack[r]{\footnotesize \textcolor{firebrick}{\textbf{OH 95\%}}\\\footnotesize O 5\%}}
\psfrag{b}[lc][lc][0.8]{~\footnotesize \textcolor{firebrick}{\textbf{OH}}}

\psfrag{c}[rc][rc][0.8]{\Shortstack[r]{\footnotesize \textcolor{firebrick}{\textbf{OH 95\%}}\\\footnotesize O 5\%}}
\psfrag{bbb}[rc][rc][0.8]{\footnotesize \textcolor{firebrick}{\textbf{OH~}}}
\psfrag{do}[cc][cc][0.8]{\footnotesize 0.5}
\psfrag{d}[rc][rc][0.8]{\footnotesize \textcolor{firebrick}{\textbf{OH 98\%}}}
\psfrag{44}[ct][ct][0.8]{\footnotesize 4.4}
\psfrag{14}[ct][ct][0.8]{\footnotesize 1.4}
\psfrag{05}[ct][ct][0.8]{\footnotesize 0.5}

\psfrag{e}[lc][lc][0.8]{\Shortstack[l]{\footnotesize \textbf{f:} 16\\\footnotesize \textbf{b:} 6.8}}
\psfrag{f}[rc][rc][0.8]{\Shortstack[r]{\footnotesize \textbf{f:} \textcolor{firebrick}{\textbf{OH 81\%}}\\\footnotesize O 9\%\\\footnotesize H 8\%\\\footnotesize \textbf{b:}  H\textsubscript{2}O 93\%}}
\psfrag{CH3l+Ml}[cb][cb][0.8]{\footnotesize CH\textsubscript{3}(+M)}
\psfrag{65}[ct][ct][0.8]{\footnotesize 6.5}
\psfrag{19}[ct][ct][0.8]{\footnotesize 1.9}
\psfrag{h}[lt][lt][0.8]{\Shortstack[l]{\footnotesize HO\textsubscript{2} 57\%\\\footnotesize O\textsubscript{2} 43\%}}
\psfrag{HO2 1Y}[rc][rc][0.8]{HO\textsubscript{2} 1\%}
\psfrag{20}[ct][ct][0.8]{\footnotesize 2.0}
\psfrag{i}[lt][lt][0.8]{\footnotesize O\textsubscript{2} 99\%}
\psfrag{kk}[lc][lc][0.8]{\Shortstack[l]{\footnotesize \textbf{f:} 4.3\\\footnotesize \textbf{b:} 2.1}}
\psfrag{k}[rc][rc][0.8]{\Shortstack[l]{\footnotesize \textbf{f:} \textcolor{firebrick}{\textbf{OH}}\\\footnotesize \textbf{b:} CH\textsubscript{3}}}
\psfrag{l}[cc][cc][0.8]{\Shortstack[l]{\footnotesize \textbf{f:} 4.4\\\footnotesize \textbf{b:} 2.2}}
\psfrag{m}[rc][rc][0.8]{\Shortstack[r]{\footnotesize \textbf{f:} CH\textsubscript{4} 98\%\\\footnotesize \textbf{b:} CH\textsubscript{3} 96\%}}

\psfrag{12}[ct][ct][0.8]{\footnotesize 1.2}
\psfrag{n}[lc][lc][0.8]{\Shortstack[l]{\footnotesize CH\textsubscript{4} 70\%\\\footnotesize H\textsubscript{2} 24\%}}

\psfrag{08}[ct][ct][0.8]{\footnotesize 0.8}
\psfrag{o}[lb][lb][0.8]{\footnotesize \textcolor{firebrick}{\textbf{OH}}}
\psfrag{p}[rc][rc][0.8]{\Shortstack[r]{\footnotesize \textbf{f:} 5.0\\\footnotesize \textbf{b:} 2.0}}
\psfrag{q}[lc][lc][0.8]{\Shortstack[r]{\footnotesize \textbf{f:} \textcolor{firebrick}{\textbf{OH 51\%}}\\\footnotesize H 33\%\\\footnotesize O 16\%\\\footnotesize \textbf{b:}  H\textsubscript{2} 73\%\\\footnotesize H\textsubscript{2}O 17\%}}

\psfrag{r}[cc][cc][0.8]{\Shortstack[l]{\footnotesize \textbf{f:} 0.8\\\footnotesize \textbf{b:} 0.6}}
\psfrag{s}[rc][rc][0.8]{\Shortstack[r]{\footnotesize \textbf{f:} CH\textsubscript{3} 59\%\\\footnotesize \textcolor{firebrick}{\textbf{OH 35\%}}\\\footnotesize O 6\%\\\footnotesize \textbf{b:}  CH\textsubscript{4} 82\%\\\footnotesize CH\textsubscript{2}O 7\%\\\footnotesize H\textsubscript{2}O 6\%}}

\psfrag{18}[ct][ct][0.8]{\footnotesize 1.8}
\psfrag{t}[rc][rc][0.8]{\footnotesize CH\textsubscript{3}(+M)}

\psfrag{u}[rc][rc][0.8]{\Shortstack[r]{\footnotesize HO\textsubscript{2} 94\%\\\footnotesize O\textsubscript{2} 6\%}}

\psfrag{05}[ct][ct][0.8]{\footnotesize 0.5}
\psfrag{v}[cb][cb][0.8]{\Shortstack[r]{\footnotesize (+M) 88\%\\\footnotesize O\textsubscript{2} 12\%}}

\psfrag{w}[cb][cb][0.8]{\footnotesize H}

\includegraphics[width=\textwidth, draft=False]{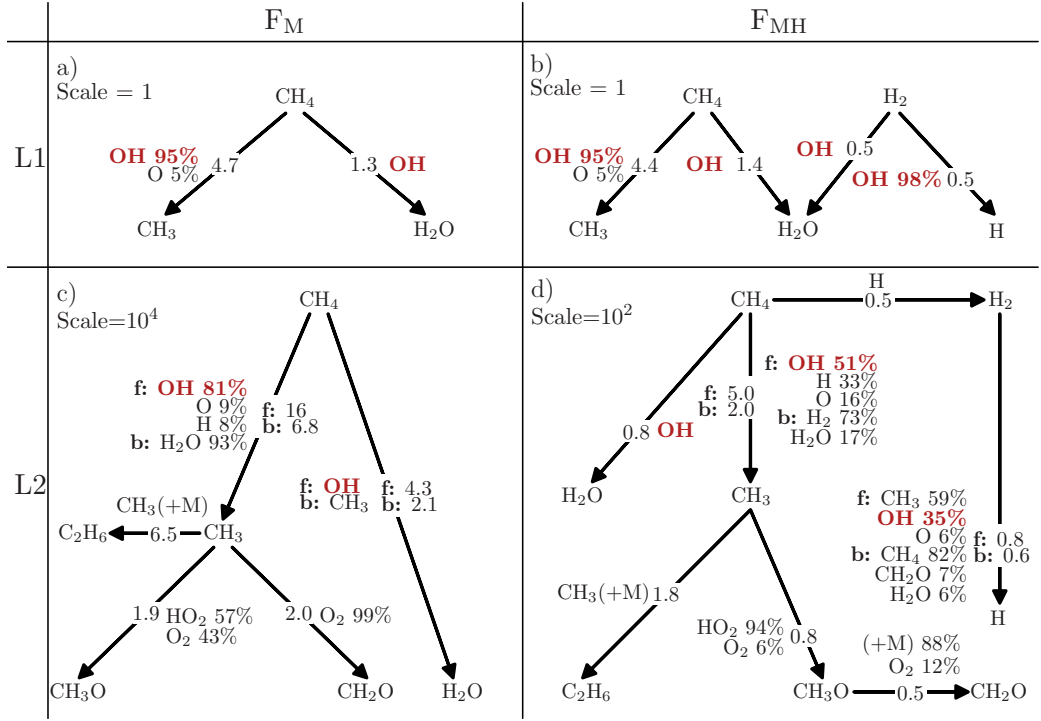}
\caption{Reaction Path Analysis (RPA) diagrams following the H atom for \opA{} (a,c) and \opB{} (b,d). Diagrams on the top (a,b) represent location L1; diagrams on the bottom (c,d) represent location L2. Fluxes units are kmol/m$^3$/s and, in each subfigure, their values are multiplied by the indicated scale factor. \textbf{f} and \textbf{b} \qm{indicate} the forward and backward fluxes, respectively. Fluxes smaller than 10\% of the maximum flux are neglected. For each flux, species contributing to less than 5\% are not represented.}
\label{fig:reactionpaths}
\end{figure*}
The different chemical behaviors at these locations are summarized in \fref{fig:reactionpaths}. For L1, the main reaction paths involve the decomposition of \qm{\methane{} to CH\textsubscript{3}, and of \hydrogen{} to H and H\textsubscript{2}O}. These paths feature high fluxes, two to four orders of magnitude higher than L2 fluxes. Reactions are mainly triggered by OH radicals coming from the combustion products of the first stage flame. Indeed, the average mass fraction of OH upstream of the fuel injector is about \mt{10^{-4}}, a significant amount that promotes the oxidation of the fuel.

Conversely, RPA diagrams for L2 highlight a change in the main fuel decomposition pathways due to the appearance of previously generated intermediate oxidized and dissociated products of fuel and to the partial consumption of the OH reservoir coming from the first stage.
Atomic H and O start playing an important role in \methane{} decomposition to CH\textsubscript{3}. Fluxes targeting CH\textsubscript{2}O become relatively important for these mixtures.

The higher fluxes of \opB{} compared to \opA{} quantify the higher reactivity of the \methane/\hydrogen{} mixture, which will ultimately lead to auto-ignition within the SB.

\subsection{Vitiated flow's relaxation towards equilibrium}
\mi{\nn{In the first stage of a sequential combustor, OH radicals are produced and flow towards the dilution air mixer.} The OH mass fraction found at the second stage fuel injector location, however, may significantly differ from the one \nn{just upstream of the dilution air mixer} due to the interaction with the cold dilution air jets. \nn{In addition to} the pure mixing effect, the large temperature drop shifts the chemical equilibrium point of the mixture. Chemical activity is therefore \nn{playing an \qm{important} role} in the vitiated flow as it evolves towards the new chemical equilibrium point \nn{at the sequential fuel injector. Moreover,} the interplay between the chemical kinetics and the residence time between the two stages determines whether this \nn{chemical relaxation} process can be completed before the second stage fuel injection.}

\begin{figure}[t!]
     \centering
     \psfrag{MM}[bc][bc][1.0]{Y\textsubscript{OH} [-]}
     \psfrag{AA}[br][br][1.0]{10\textsuperscript{-6}}
     \psfrag{BB}[br][br][1.0]{10\textsuperscript{-5}}
     \psfrag{CC}[br][br][1.0]{10\textsuperscript{-4}}
     \psfrag{GGG}[cc][cc][1.0]{1200}
     \psfrag{HHH}[cc][cc][1.0]{1350}
     \psfrag{MMM}[cc][cc][1.0]{1500}
     \psfrag{VVV}[cc][cc][1.0]{Temperature [K]}
     \psfrag{U}[][][1.0]{0}
     \psfrag{EE}[][][1.0]{0.1}
     \psfrag{FF}[][][1.0]{0.2}
     \psfrag{KK}[][][1.0]{\Coh}
     \psfrag{HH}[][][1.0]{$C=\frac{Y_{OH-Eq.} \big\rvert_{T,P}}{Y_{OH}}$}
     \psfrag{Yin}[][][1.0]{Y\mt{_{OH}^{inlet}}}
     \psfrag{Yequ}[tl][tl][1.0]{Y\mt{_{OH}^{eq.}}}
     \includegraphics[width=0.45\textwidth, draft=False]{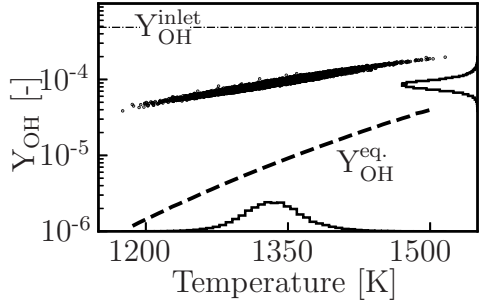}
 \caption{Scatter plot representing the mass fraction of OH as function of the temperature on a plane right upstream of the fuel injector. Points distribution is projected on both axes.}
 \label{fig:OHequi}
 \end{figure}

\mi{Given the importance of OH on the ignition \nn{chemistry}, its mass fraction ($\mathrm{Y}_{\mathrm{OH}}$) is taken as reference to quantify the distance of the advected mixture from the chemical equilibrium point. Individual samples of $\mathrm{Y}_{\mathrm{OH}}$ are extracted on a plane right upstream of the fuel injector from multiple instantaneous LES solutions and plotted as function of the temperature in \fref{fig:OHequi}. The $\mathrm{Y}_{\mathrm{OH}}$ points distribution is compared with two important lines, namely:
\begin{itemize}
    \item[-] the frozen chemistry line $\mathrm{Y}_{\mathrm{OH}}^{\mathrm{inlet}}$, which represents the mass fraction of OH injected at the inlet of the domain, \nn{with the assumption that the residence time in the first stage chamber (not simulated in this work) is long enough for having reached chemical equilibrium at 1860~K;}
    \item[-] the equilibrium line $\mathrm{Y}_{\mathrm{OH}}^{\mathrm{eq.}}$, which represents the equilibrium mass fraction of OH as function of the temperature, reached in case of fast enough chemistry \nn{once the inlet flow and the dilution air have perfectly mixed.}
\end{itemize}}
\mi{In \fref{fig:OHequi}, the LES data highlights that the equilibrium is not reached: there is still margin for the OH mass fraction to decrease. $\mathrm{Y}_{\mathrm{OH}}$ can span more than two orders of magnitude between $\mathrm{Y}_{\mathrm{OH}}^{\mathrm{inlet}}$ and $\mathrm{Y}_{\mathrm{OH}}^{\mathrm{eq.}}$, suggesting that the combustion behavior of the sequential combustor can be strongly affected by the distance of the vitiated \qm{air} from chemical equilibrium, as will be confirmed in \sref{sec:equilibriumeffect}. Parameters influencing the relaxation towards equilibrium \nn{just upstream of the sequential fuel injection} become therefore crucial for the operability of this type of combustors, including not only the thermodynamic properties of the system (i.e. the operating point), but also the combustor geometry, which directly influences the residence time \nn{and dilution air mixing process} between the two stages.}

\subsection{\mi{Influence of equilibrium on the ignition process}}
\label{sec:equilibriumeffect}
\mi{The effect of chemical equilibrium on the ignition behavior of the sequential stage is investigated via an additional LES labelled \opC{}. 
\opC{} has the same simulation parameters as \opB{}, except for two numerical artifacts:
\begin{itemize}
    \item[-] the chemical reactions are disabled between the inlet and the leading edge of the fuel injector;
    \item[-] \qm{the mixture composition at the inlet is adjusted so that, after mixing with the dilution air, the vitiated air flow is at equilibrium just upstream of the fuel injector.}
\end{itemize}
This strategy allows to artificially construct the limit case where the \nn{air-diluted} vitiated flow has, on average, enough time to \nn{reach equilibrium composition} before the second stage fuel injector. Temperature stratification due to the imperfect mixing with the dilution air is preserved, thereby allowing to isolate the effects of chemical equilibrium. \opB{} and \opC{} are compared in terms of chemical activity and CEMA \qm{quantities} in Secs. \ref{sec:chem_act} and \ref{sec:CEMA_fmh}.}
\subsubsection{Chemical Activity in the Sequential Burner}\label{sec:chem_act}
\begin{figure}[h!]
     \psfrag{a}[lb][lb][1.0]{a)}
     \psfrag{b}[lb][lb][1.0]{b)}
     \psfrag{00m}[][][1.0]{\small 0.0 ms}
     \psfrag{02m}[][][1.0]{\small 0.2 ms}
     \psfrag{04m}[][][1.0]{\small 0.4 ms}
     \psfrag{06m}[][][1.0]{\small 0.6 ms}
     \psfrag{08m}[][][1.0]{\small 0.8 ms}
     \psfrag{HRR}[bc][bc][1.0]{$\dot{Q}$ [W/m\textsuperscript{3}]}
     \psfrag{1e7}[bc][bc][1.0]{10\textsuperscript{7}}
     \psfrag{1e8}[bc][bc][1.0]{10\textsuperscript{8}}
     \psfrag{1e9}[bc][bc][1.0]{10\textsuperscript{9}}
     \psfrag{HR}[cc][ct][1.0]{$\overline{\dot{\mathcal{Q}}}(z)$ [MW/m\textsuperscript{3}]}
     \psfrag{Coord}[][][1.0]{}
     \psfrag{0}[rc][rc][1.0]{0}
     \psfrag{4}[rc][rc][1.0]{40}
     \psfrag{8}[rc][rc][1.0]{80}
     \psfrag{FI}[][][1.0]{Fuel Inj.}
     \psfrag{CC}[][][1.0]{Comb. C.}
     \psfrag{Fmh}[lt][lt][1.0]{\opB}
     \psfrag{FMH}[lb][lb][1.0]{\opB}
     \psfrag{Fmhs}[lt][lt][1.0]{\opC}
     \psfrag{FMHS}[rb][rb][1.0]{\opC}
     \psfrag{0p}[lc][lc][1.0]{\textcolor{cornflowerblue}{0\%}}
     \psfrag{1p}[lc][lc][1.0]{\textcolor{cornflowerblue}{10\%}}
     \psfrag{2p}[lc][lc][1.0]{\textcolor{cornflowerblue}{20\%}}
     \psfrag{HRint}[cc][cc][1.0]{\textcolor{cornflowerblue}{$\overline{\mathcal{P}}_{\%}(z)$ [-]}}
     \psfrag{z}[][][1.0]{$z$}
     \psfrag{y}[][][1.0]{$y$}
     \centering
     \includegraphics[width=\textwidth, draft=False]{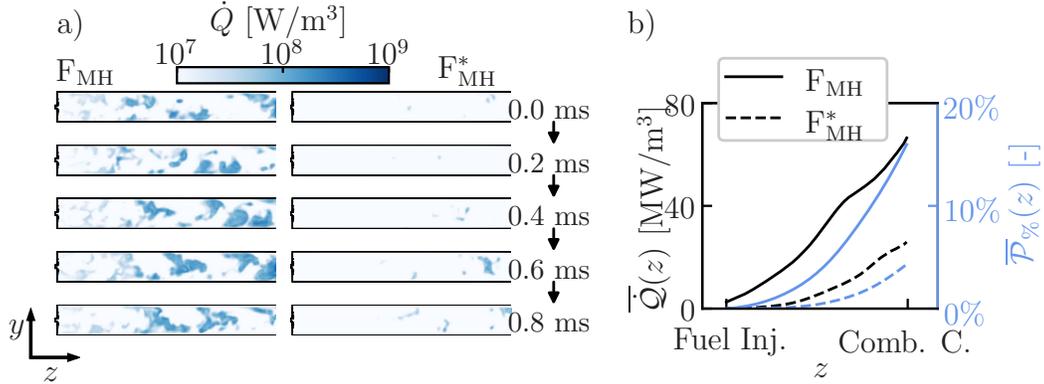}
  \caption{\qm{Planar cuts of the sequential burner colored by $\dot{Q}$ for both \opB{} (left column) and \opC{} (right column) for multiple instants spaced by 0.2 ms (a), and profiles of $\overline{\dot{\mathcal{Q}}}(z)$ and $\overline{\mathcal{P}}_{\%}(z)$ in the sequential burner as function of the streamwise coordinate (b).}}
 \label{fig:hr_fmh_fmhs}
 \end{figure}

\mi{\Fref{fig:hr_fmh_fmhs}a highlights significant differences between \opB{} and \opC{} in terms of auto-ignition kernels generation in the SB. While \opB{} features an intense production of auto-ignition kernels in multiple regions of the SB, \opC{} is characterized by much weaker auto-ignition kernel formation, limited to some sporadic events towards the end of the SB. \qm{In \fref{fig:hr_fmh_fmhs}b, two quantities are represented as function of the streamwise coordinate $z$: the time-averaged integral mean of the heat release rate $\overline{\dot{\mathcal{Q}}}(z)$, and the time-averaged normalized total power released up to $z$  $\overline{\mathcal{P}}_{\%}(z)$.}
\qm{These quantities are defined as follows:
\begin{equation}
    \overline{\dot{\mathcal{Q}}}(z)= \int_0^{\Delta t}\frac{ \iint \dot{Q}(x,y,z,t)\,dx\,dy }{A(z)\Delta t}dt
\end{equation}
\begin{equation}
\overline{\mathcal{P}}_{\%}(z)=\int_0^{\Delta t}\frac{\int_{inlet}^z \iint \dot{Q}(x,y,z^*,t)\,dx\,dy\,dz^* }{P_{2nd}^{th.}\Delta t}dt
\end{equation}
where $\dot{Q}(x,y,z)$ is the heat release rate field, $A(z)$ is the area of the cross-section normal to the $z$ axis at location $z$, $P_{2nd}^{th.}$ is the thermal power of the second stage, and $\Delta t${=10~ms}.
$\overline{\dot{\mathcal{Q}}}(z)$ and $\overline{\mathcal{P}}_{\%}(z)$ confirm} the trends depicted in the 2D fields on a longer timescale and on the whole 3D domain, highlighting that the release of heat from the mixture happens right after the fuel injector in \opB{}, while it is postponed towards the end of the SB for \opC{}. The final value reached by $\overline{\mathcal{P}}_{\%}(z)$ quantifies the power released, on average, in the SB: approximately 16\% of the total fuel power is released by means of auto-igniting kernels in \opB{}, while this percentage drops to approximately 4\% in \opC{}.}

\begin{figure}[h!]
     \psfrag{aa}[lb][lb][1.0]{a)}
     \psfrag{bb}[lb][lb][1.0]{b)}
     \psfrag{cc}[lb][lb][1.0]{c)}
     \psfrag{dd}[lb][lb][1.0]{d)}
     
     \psfrag{wCH4}[bc][bc][1.0]{\RateM{} [mol/m\textsuperscript{3}/s]}
     \psfrag{frac75}[bc][bc][1.0]{\PercM{} [-]}
     \psfrag{wH2}[bc][bc][1.0]{\RateH{} [mol/m\textsuperscript{3}/s]}
     \psfrag{frac7}[bc][bc][1.0]{\PercH{} [-]}
     \psfrag{80}[bc][bc][1.0]{80\%}
     \psfrag{65}[bc][bc][1.0]{65\%}
     \psfrag{50}[bc][bc][1.0]{50\%}
     \psfrag{100}[bc][bc][1.0]{$\mathrm{100}$}
     \psfrag{10}[bc][bc][1.0]{$\mathrm{10}$}

     \psfrag{FMH}[lb][lb][1.0]{\opB}
     \psfrag{FMHS}[lt][lt][1.0]{\opC}
     
     \psfrag{z}[][][1.0]{$z$}
     \psfrag{y}[][][1.0]{$y$}
     \centering
     \includegraphics[width=\textwidth, draft=False]{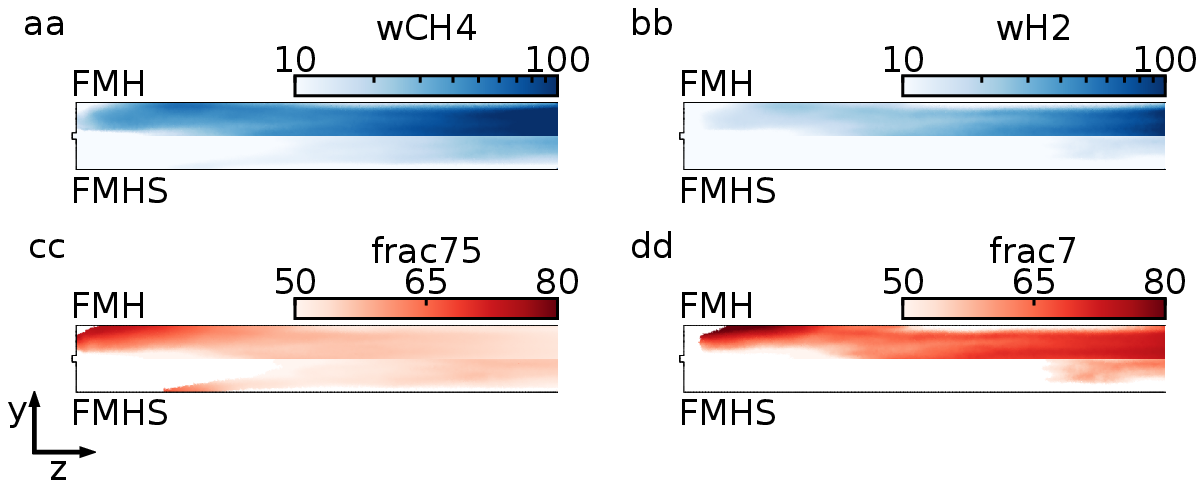}
 \caption{Planar cuts of the SB colored by the \methane{} decomposition rate \RateM{} (a), the \hydrogen{} decomposition rate \RateH{} (b), \PercM{} (c), and \PercH{} (d). \PercM{} and \PercH{} are represented only in the regions where \RateM{} and \RateH{} are above the threshold value of 10 mol/m\textsuperscript{3}/s. The fields are the result of a 10 ms average.}
 \label{fig:analysis_fmh_fmhs}
 \end{figure}

\qm{\Fref{fig:analysis_fmh_fmhs} depicts the time-averaged consumption rates of \methane{} (\RateM{}) and \hydrogen{} (\RateH{}) in the SB.} \mi{They are computed summing the contributions of all the reactions characterized by negative stoichiometric coefficients for the fuel molecules. Reversible reactions are split into forward and backward reactions that contribute individually to the sums. \PercM{} and \PercH{} quantify the relative contributions of reactions \RM{} (\ce{CH4 + OH -> CH3 + H2O}) and \RH{} (\ce{H2 + OH -> H + H2O}) to \RateM{} and \RateH{}, respectively. \qm{The averaging operation is performed over 10~ms.}}

\mi{In \opB{}, the out-of-equilibrium flow composition promotes the formation of a chemically active region in the recirculation zone attached to the fuel injector \qm{(Z1 in \fref{fig:TV})} where both \methane{} and \hydrogen{} molecules are decomposed. In agreement with the analysis in \sref{sec:ignitionmechanism}, OH is the main responsible for fuel consumption with the reactions \RM{} and \RH{} accounting for more than 80\% of \RateM{} and \RateH{}, respectively. Proceeding downstream in the SB, part of the excess OH advected from the first stage is consumed and a pool of active radicals and intermediate species is progressively built: new pathways of \methane{} decomposition are available and \PercM{} decreases to roughly 50\%. The \hydrogen{} consumption pathways are less affected by this behavior and \PercH{} remains above 70\% proceeding towards the end of the SB.}

\mi{Conversely, in \opC{}, the mixture is almost chemically inert in the SB: no fuel decomposition is taking place close to the fuel injector and both \RateM{} and \RateH{} are approximately one order of magnitude smaller than in \opB{}. Furthermore, the relative importance of OH is reduced, as \PercM{} and \PercH{} are almost constant around 50\%.}

\subsubsection{\nn{Chemical Explosive Mode Analysis}}\label{sec:CEMA_fmh}
\begin{figure}[h!]
     \psfrag{aa}[lb][lb][1.0]{a)}
     \psfrag{bb}[lb][lb][1.0]{b)}
     \psfrag{cc}[lb][lb][1.0]{c)}
     \psfrag{dd}[lb][lb][1.0]{d)}
     \psfrag{FMH}[lb][lb][1.0]{\opB{}}
     \psfrag{FMHS}[lb][lb][1.0]{\opC{}}

     \psfrag{alpha}[cc][cc][1.0]{$\alpha$}
     \psfrag{EI}[cc][cc][1.0]{EI}
    \psfrag{AA}[t][t][1.0]{\mt{-\infty}}
    \psfrag{BB}[t][t][1.0]{\mt{-1}}
    \psfrag{CC}[t][t][1.0]{\mt{1}}
    \psfrag{DD}[t][t][1.0]{\mt{+\infty}}
    \psfrag{EEE}[t][t][1.0]{\small T}
    \psfrag{FFF}[t][t][1.0]{\small CH\textsubscript{3}}
    \psfrag{GGG}[t][t][1.0]{\small CH\textsubscript{2}O}
    \psfrag{MMM}[t][t][1.0]{\small Other}
    \psfrag{time}[][][1.0]{Time [ms]}
    \psfrag{lam}[][][1.0]{$\log_{10}(1+\lambda)$}
    \psfrag{0}[][][1.0]{0}
    \psfrag{2}[][][1.0]{2}
    \psfrag{4}[][][1.0]{4}
    \psfrag{0m}[][][1.0]{0}
    \psfrag{2m}[][][1.0]{2}
    \psfrag{4m}[][][1.0]{4}
    \psfrag{6m}[][][1.0]{6}
    \psfrag{0t}[][][1.0]{0}
    \psfrag{0.5t}[][][1.0]{0.5}
    \psfrag{1t}[][][1.0]{1}
    \psfrag{0.1}[][][1.0]{0.1}
    \psfrag{0.3}[][][1.0]{0.3}
    \psfrag{0.6}[][][1.0]{0.6}
    \psfrag{0.2}[][][1.0]{0.2}
    \psfrag{0.7}[][][1.0]{0.7}
    \psfrag{ei}[][][1.0]{EI [-]}
    \psfrag{omn}[lc][lc][1.0]{\ReacB{}}
    \psfrag{nom}[lc][lc][1.0]{\ReacC{}}
    \psfrag{onm}[lc][lc][1.0]{\ReacD{}}
    \psfrag{ymh}[rc][rc][1.0]{\ReacB{}}
    \psfrag{ymhs}[rc][rc][1.0]{\ReacC{}}
    \psfrag{ymhz}[rc][rc][1.0]{\ReacD{}}
    \psfrag{CH2O}[cc][cc][1.0]{\textcolor{turquoise}{CH\textsubscript{2}O}}
    \psfrag{CH3}[cc][cc][1.0]{\textcolor{purple}{CH\textsubscript{3}}}
    \psfrag{T}[cc][cc][1.0]{\textcolor{orange}{T}}
    \psfrag{y}[cc][cc][1.0]{$y$}
    \psfrag{z}[cc][cc][1.0]{$z$}
    \psfrag{tau}[][][1.0]{$\mathrm{Time}/\tau_{AI}$ [-]}

     \centering
     \includegraphics[width=\textwidth, draft=False]{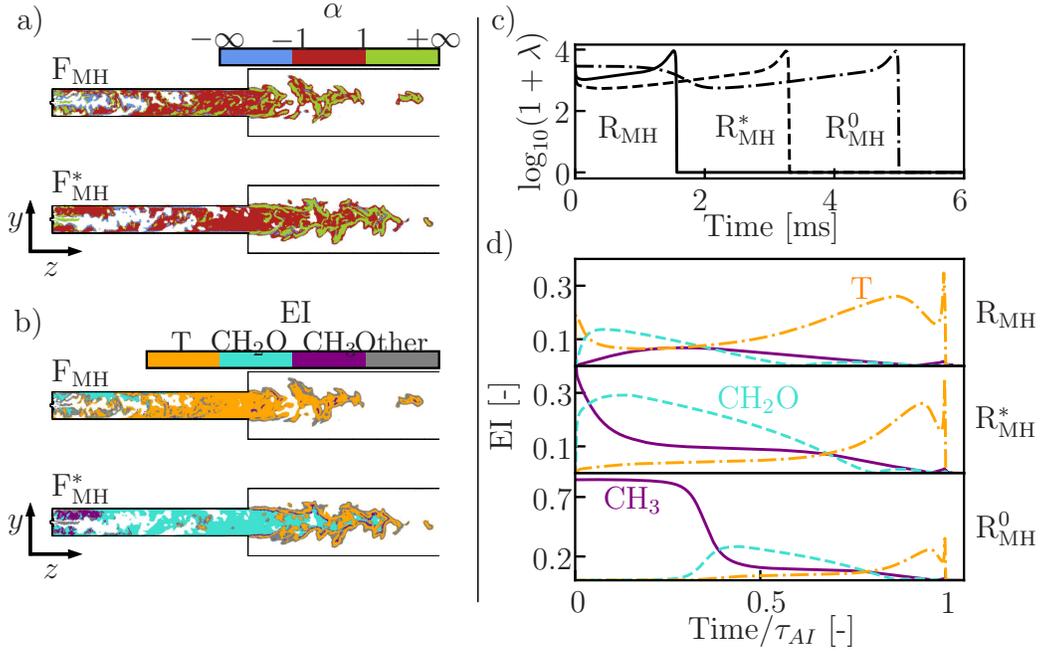}
 \caption{Planar cuts colored by $\mathrm{\alpha}$ (a) and \textbf{EI} largest entry (b) for \opB{} and \opC{} (left). Results of CEMA for 0D reactors in terms of eigenvalues $\lambda_e$ (c) and Explosive Indexes of temperature, CH\textsubscript{2}O and CH\textsubscript{3} (d) (right).}
 \label{fig:CEMA_fmh_fmhs}
 \end{figure}

\mi{\Fref{fig:CEMA_fmh_fmhs} depicts the results of CEMA applied to \opB{}, \opC{}, and to 0D reactor simulations computed with Cantera \qm{\cite{cantera}}. Three 0D simulations are carried out, namely \ReacB{}, \ReacC{}, and \ReacD{}. \qm{\ReacB{} and \ReacC{}} are designed to mimic the 3D LES cases \opB{} and \opC{}, respectively, while \ReacD{} represents a limit case where the vitiated flow consists only of major species (i.e. N\textsubscript{2}, O\textsubscript{2}, H\textsubscript{2}O, and CO\textsubscript{2}). The 0D reactors are adiabatic and isobaric, and the initial state is obtained by mixing fuel and vitiated air flow. \qm{The fuel composition} is fixed and is representative of the fuel \qm{blend} of the 3D LES; the \qm{vitiated air composition} is properly varied according to the specific case investigated.}

\mi{Only the regions where $\lambda_e \geq 100~s^{-1}$ are represented in the planar cuts of \fref{fig:CEMA_fmh_fmhs}a and \ref{fig:CEMA_fmh_fmhs}b, indirectly highlighting how $\lambda_e$ is above the threshold value in large portions of the domain also in \opC{}. This is expected also from the 0D analysis, where the vitiated flow composition has a limited influence on the $\lambda_e$ value.}

\mi{The extended $\lvert \alpha \rvert < 1$ regions upstream of the flame front depict a propagating flame regime supported by auto-ignition chemistry in \opC{}. The vitiated flow composition has therefore no effect on the predicted flame regime. This finding does not contradict the already discussed suppression of auto-ignition kernels in \opC{}. Indeed, $\alpha$ refers to the asymptotic behavior of the mixture, but the capability of forming auto-ignition kernels within the physical boundaries of the system depends also on the competition between the residence time and the auto-ignition delay time. While the \qm{residence time} is almost constant and corresponds to the convective time between fuel injector and sequential flame, the \qm{auto-ignition delay} is influenced by the composition of the vitiated air, as can be seen from \fref{fig:CEMA_fmh_fmhs}c. Net of stratification and turbulence effects that are not captured in 0D reactors, the larger amount of \qm{radicals} in the vitiated air, the faster the ignition.}

\mi{Conversely, the dominating explosive mode is significantly affected by the minor species in the vitiated air, as quantified by the largest entry of the explosive index vector $\mathbf{EI}$. In \opB{}, as highlighted already in \sref{sec:CEMA}, the explosive index associated to temperature is dominating almost everywhere, with CH\textsubscript{2}O dominating close to the fuel injector. In \opC{}, on the other hand, the largest entry of the explosive index vector is mainly CH\textsubscript{2}O, except for the regions close to the fuel injector and close to the flame front, dominated by CH\textsubscript{3} and temperature, respectively. These trends are explained with the help of \fref{fig:CEMA_fmh_fmhs}d, which highlights two phases of the methane/hydrogen blend ignition: a fuel decomposition phase, with the dominant components of $\mathbf{EI}$ being CH\textsubscript{3} and CH\textsubscript{2}O; and a thermal runaway phase, with $\mathbf{EI}$ dominated by temperature. The time spent in these phases is influenced by the concentration of OH in the initial mixture. In particular, the higher the OH concentration, the faster the fuel decomposition. In \opC{}, the OH radicals reaching the second stage shorten the CH\textsubscript{3} dominated phase but are not enough to trigger the thermal runaway phase, consequently leading to almost no auto-ignition kernel formation in the SB.}

\section{Conclusions}
\mi{The present paper reports the results of a numerical investigation on the ignition and combustion behavior of pure \methane{} fuel (\qm{case} \opA{}) and of a \nn{\hydrogen{}-enriched \methane{}} fuel (\qm{case} \opB{}) in the second stage of a sequential combustor, performed using LES in combination with a precise description of the chemistry. 
It is found that, \nn{for the operating conditions considered,} replacing 4\% in mass of \methane{} with \hydrogen{} significantly changes the behavior of the sequential flame, causing auto-ignition events \qm{in the sequential burner, upstream of the sequential flame}.
Chemical Explosive Mode Analysis is used to analyze the combustion regimes of the two \nn{cases, each of them featuring the same temperature and mass flow of both the vitiated flow and the dilution air}, highlighting that \opA{} features a purely propagating flame in the combustion chamber, while \opB{} is characterized by a mixed combustion regime with a propagating flame supported by upstream auto-ignition chemistry. 
The \nn{combustion modes in the sequential burner and chamber are} strongly influenced by minor combustion products advected from the first stage, especially OH radicals. Indeed, as shown via Reaction Path Analysis \qm{(RPA)}, they induce a prompt fuel oxidation near the fuel injector, enhancing the chemical activity in the sequential burner and ultimately leading to the formation of auto-ignition kernels in \qm{the \opB{} case}.
This is primarily due to the large \qm{concentration} of OH upstream of the fuel injector, whose mass fraction is, \qm{on average, approximately an order of magnitude higher than the one at the theoretical chemical equilibrium reached assuming perfect mixing between the vitiated and the dilution air}. The importance of the out-of-equilibrium composition in determining the combustion mode of the second stage is demonstrated performing an additional LES (\opC{}) characterized by \qm{the vitiated air flow at chemical equilibrium upstream of the fuel injector}. Compared to \opB{}, \opC{} features no prompt fuel oxidation in the sequential burner and \qm{the} inhibition of auto-ignition kernels formation.
Therefore, the parameters affecting the relaxation towards chemical equilibrium of the vitiated flow are expected to interfere with the behavior of sequential combustors operating at auto-ignition conditions with varying fractions of \hydrogen{} blending. These include both the thermodynamic conditions, which define the equilibrium point, and the combustor geometry, which affects the residence time and determines whether the equilibrium can be reached.}


\section*{Acknowledgements}
This project has received funding from the European Research Council (ERC) under the European Union’s Horizon 2020 research and innovation program (grant agreement No [820091]).
We acknowledge PRACE for awarding us access to MareNostrum at the Barcelona Supercomputing Center (BSC), Spain (project ID 2021250004).
This work was supported by a grant from the Swiss National Supercomputing Centre (CSCS) under project ID s1138.
The authors thank P. Pepiot for providing the YARC chemistry reduction tool.
They also gratefully acknowledge CERFACS for providing the LES solver AVBP; they especially thank G. Staffelbach and O. Vermorel for the technical support.

\bibliographystyle{elsarticle-num}
\bibliography{bibliography}

\end{document}